\pdfoutput=1
  \documentclass[manuscript]{aastex}	
\usepackage{graphicx,color}		
\usepackage{amssymb}			
\usepackage{color}			
\usepackage{url}			
\usepackage{amsmath}			
\usepackage{rotating}			
\usepackage{float}			
\usepackage{textcomp}			
\usepackage{psfig}
\usepackage{dcolumn}
\usepackage{times}
\usepackage{tabularx}
\usepackage[colorlinks=true,citecolor=blue]{hyperref}
\usepackage{subfig}
\usepackage{ragged2e}

\begin{document}
\title{Variation of Supergranule Parameters With Solar Cycles: Results From Century-Long Kodaikanal Digitized C\MakeLowercase{a}  $\scriptsize{{\textrm{II}}}$ K Data}
\author{Subhamoy Chatterjee$^{1}$,
Sudip Mandal$^{1}$,
Dipankar Banerjee$^{1, 2}$
}
 \affil{$^{1}$Indian Institute of Astrophysics, Koramangala, Bangalore 560034, India. e-mail: {\color{blue}{dipu@iiap.res.in}}\\
$^{2}$Center of Excellence in Space Sciences India, IISER Kolkata, Mohanpur 741246, West Bengal, India  \\}

\justify
\begin{abstract}

The century-long (1907-2007) C\MakeLowercase{a}  $\scriptsize{{\textrm{II}}}$ K spectroheliograms from Kodaikanal Solar Observatory, India, has recently been digitised and calibrated. Applying a fully-automated algorithm (which includes contrast enhancement and `Watershed method') on this data, we have identified the supergranules and calculated the associated parameters, such as scale, circularity, fractal dimension. We have segregated the quiet and active regions and obtained the supergranule parameters separately for these two domains. In this way, we have isolated the effect of large and small scale magnetic fields on such structures and find significantly different behavior of the supergranule parameters over solar cycles. Such differences indicate the intrinsic changes in the physical mechanism behind generation and evolution of supergranules in presence of small and large scale magnetic fields. This also highlights  the need for further studies using solar dynamo theory along with magneto-convection models.
\end{abstract}


\section{Introduction}

Sun, a magnetically active star, has an atmosphere which varies widely with height in density and temperature. The lower most layer of the atmosphere is known as photopshere where we observe the signature of solar convection, the granules (characteristic scale of the granules is few $\mathrm{Mm}$). There are patterns on the solar surface which are larger ($\sim$30 $\mathrm{Mm}$) than the granular scales. These are known as supergranules. Supergranules appear with different shapes and sizes. The boundaries of these structures are the host of the magnetic fluxes and dispersal of such flux is believed to be governed by the diffusion caused by evolution of supergranules \citep{{0004-637X-597-2-1200},{0004-637X-662-1-715}}.

The origin of the supergranulation is not fully understood. It can be due to convection (like granules) or can be a dynamic instability also. The study of the supergranules is important from the fact that it reveals the intrinsic scale of physical mechanism which drives these structures. Also one can investigate the effect of strong magnetic field by studying the supergranular properties separately for active and quite regions. There have been a few studies in the past on the detection and calculation of different properties of the supergranules. Using a auto-correlation curve technique \citet{{1970SoPh...13..292S},{0004-637X-481-2-988}} have detected the supergranular structures whereas \citet{1989A&A...213..431M} have estimated the latitudinal dependence of the supergranule sizes, by using C\MakeLowercase{a}  $\scriptsize{{\textrm{II}}}$ images, through a Fast Fourier Transform (FFT) analysis for both the active and quiet regions. These automated methods are useful in determining the aggregate properties of supergranules but they fail to record the parameters of individual structures. In an another work, \citet{1999A&A...344..965B} have used an automated skeleton detection method and have shown the temporal variation of the quiet region region cell size for one year data set. \citet{0004-637X-481-2-988} showed the invariance of supergranule scale distribution at different spatial smoothing. They have also studied the distribution of supergranule scales and found similarity of the same with Voronoi tessellation. A similar work has been done by \citet{0004-637X-534-2-1008} where these authors have quantified the distribution using skewness and kurtosis parameters. In a recent work, \citet{2041-8205-730-1-L3} have explored the variation of the supergranular parameters from five independent sources using the `watershed segmentation' method to detect the supergranules from the images. From their results (for a period of 33 years, 1944-1976) they found imprints of the solar cycle variation in the supergranule parameters. Thus we see that a study of the supergranule parameters for several solar cycles is necessary to find out their relation with the large scale solar variation. 

Using the Kodaikanal digitized C\MakeLowercase{a}  $\scriptsize{{\textrm{II}}}$ K data, we present, for the first time, the variation of the supergranular parameters for a period of 100 years (1907-2007) in this paper . Kodaikanal Solar Observatory (KSO) has archived 100 years (cycle 14-23) of chromospheric images in C\MakeLowercase{a}  $\scriptsize{{\textrm{II}}}$ K (3933.67 \AA) through an unchanged \emph{f/21} optics with a 30 cm objective.  This gives an enormous opportunity to study the synoptic variation of supergranular cell sizes over many solar cycles and also to understand their correspondence with the solar activity. This paper is organized as follows: After presenting a brief data description in Section~\ref{sec:data}, we define the detection method and various parameters associated with supergranlues in Section~\ref{sec:def}. Results from the KSO are discussed in Section~\ref{sec:kresult} whereas in Section~\ref{sec:presult} we present the results using the data from other observatories. Finally, Section~\ref{sec:summary} provides brief summary and conclusion.


\section{Data Description} \label{sec:data}
   The primary data used in this study are taken from the digitized archive of the Kodaikanal C\MakeLowercase{a} $\scriptsize{{\textrm{II}}}$ K observations \footnote{\url{https://kso.iiap.res.in/data}} \citep{{2014SoPh..289..137P},{0004-637X-827-1-87}}. Apart from that we have also used data from Precision Solar Photometric Telescope (PSPT) \citep{Ermolli1998}. The details of each data set are given in Table~\ref{table:data}.

      We have used, for comparison, the V2.0 daily sunspot number data from SIDC (Solar Influences Data Analysis Center) which is available at \url{http://www.sidc.be/silso/datafiles}.

 \begin{table}[!htbp]
\begin{center}
\centering
\caption{  Details of the different datasets used in this study}  

\label{table:data}
\begin{tabular}{lcccc r@{   }l c}

  \hline
  & \multicolumn{1}{c}{Data Source}& \multicolumn{1}{c}{Wavelngth (passband) (\r{A})} & \multicolumn{1}{c}{Duration} & \multicolumn{1}{c}{Pixel Scale}  \\
     
     \hline
     &  KSO C\MakeLowercase{a} $\scriptsize{{\textrm{II}}}$ K & 3933.67 (0.5) & 1907-2007 &  0.8" \\
     &  PSPT-ITALY & 3934.00 (2.5) &  1996-2016 & 1''\\
     &  PSPT-USA & 3934.00 (1.0) &  2005-2015 & 1''\\
      \hline

\end{tabular}
\end{center}
\end{table}

\section{Definitions And Methods} \label{sec:def}

   \subsection{Detection of Supergranules}
\begin{figure*}[!htbp]
\centering
 \captionsetup[subfigure]{labelformat=empty}
  \subfloat[ ]{\includegraphics[trim = 0mm 0mm 0mm 0mm, clip,width=0.5\textwidth]{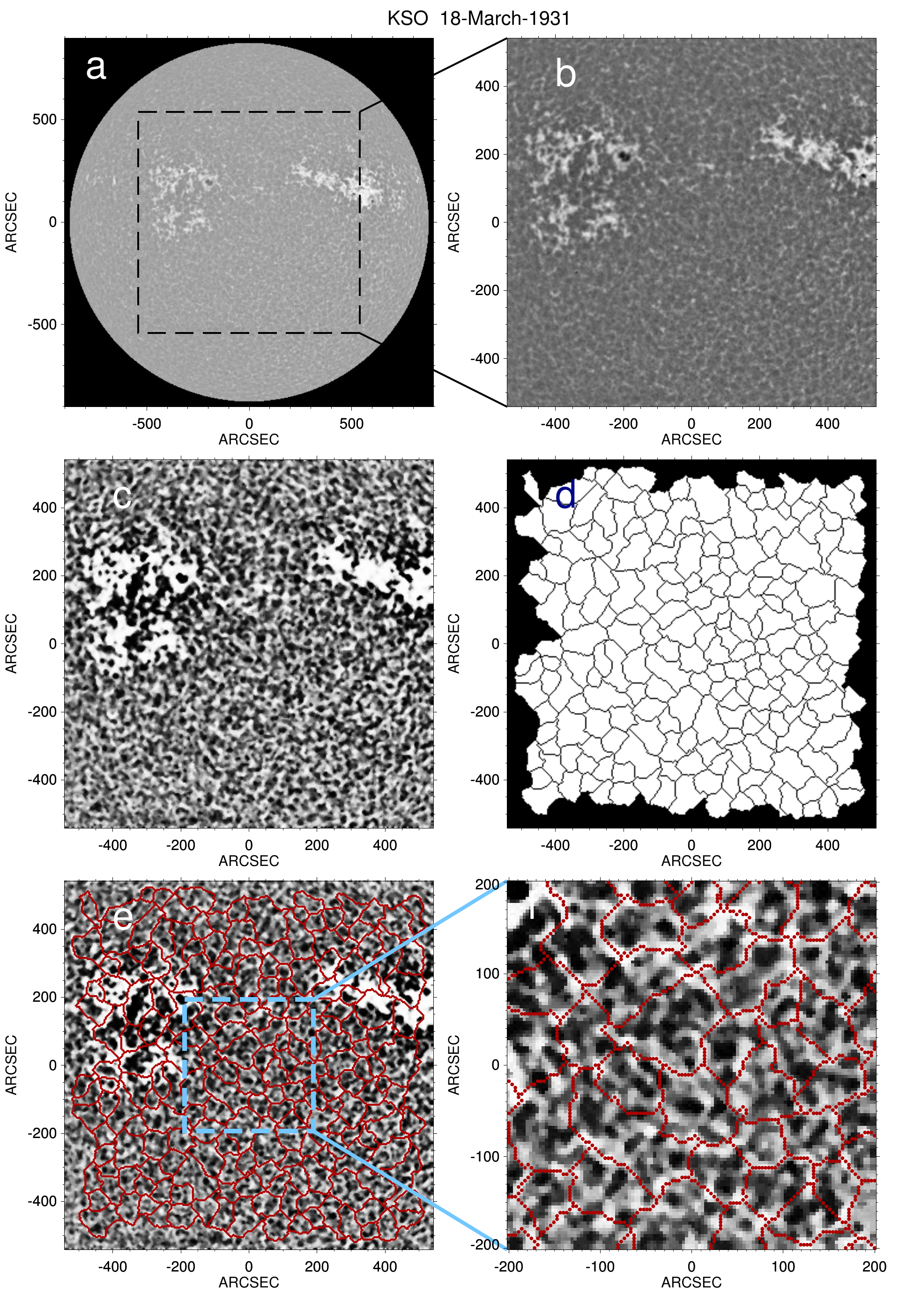}}
  \subfloat[ ]{\includegraphics[trim = 0mm 0mm 0mm 0mm, clip,width=0.5\textwidth]{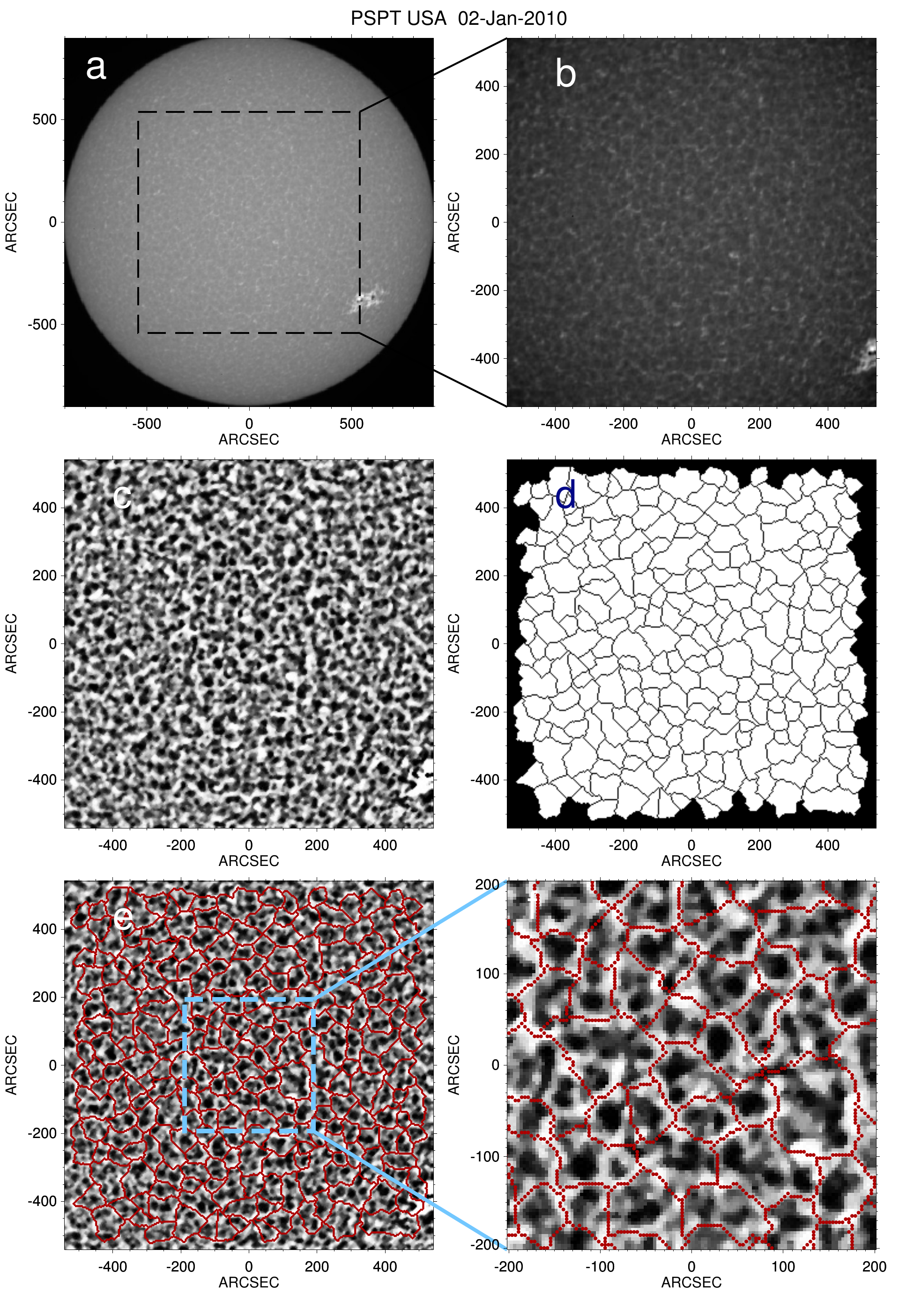}}
  \caption{Different processing steps to detect supergranules. a) Ca~{\sc ii}~K disc centered image; b) A central window [as highlighted by the rectangle in (a)] used for further processing; c) Limb darkening corrected, intensity enhanced and smoothed version of (b); d) Detected supergranules using the Watershed transform [supergranule boundaries are shown in black]; e) Supergranule boundaries from (d) overplotted on (c); f) Magnified view of panel (e) having FoV of 400"x400".}
 \label{fig:supergranule_proc}
\end{figure*}

              Supergranules refer to the large scale velocity structures with a spatial extent of $\sim$30 $\mathrm{Mm}$ and a typical lifetime of 25 hours \citep{Rieutord2010}. These structures also have a strong horizontal flow of $\approx$ 400 m s$^{-1}$. Now, in the intensity images taken through a chromospheric line such as C\MakeLowercase{a} $\scriptsize{{\textrm{II}}}$ K, we notice a boundary like pattern also known as `chromospheric network'. These networks actually outline the supergranular cells and can be used as a proxy for supergranular shape and size measurements \citep{1964ApJ...140.1120S}. We follow this convention thoughout the paper. In order to identify these structures, we first select a rectangular region [`region of interest', ROI] at the disc center with sides equal to $60\%$ of solar disc diameter. Such selection is made in order to minimize the errors due to the projection effect. Different steps, involved in detecting the supergranules from KSO and PSPT intensity images, are shown in different panels of Figure~\ref{fig:supergranule_proc}. We highlight the ROIs with the rectangles as shown in panels ~\ref{fig:supergranule_proc}.a's and the full view of the ROIs are shown in panels ~\ref{fig:supergranule_proc}.b's. Next, these regions were histogram equalized and smoothed with a median filter to reduce noise (panels~\ref{fig:supergranule_proc}.c's).

 We have used the morphological closing and subsequently Watershed transform \citep{{Vincent:1991:WDS:116700.116707},{CYTO:CYTO10079},2041-8205-730-1-L3} on those ROI images to detect the supergranules. The basic principle behind the watershed transformation is as follows: If an image is visualized as a topographic surface with gray-levels as heights, watershed segmentation divides the same into catchment basins. All pixels corresponding to a basin is connected to a local minimum falling within the same through a pixel-path of steadily decreasing intensity height \citep{sonka2014image}. Now, the rationale behind the usage of morphological closing was to avoid over segmentation into smaller scales through watershed method. The results of the watershed transformation on the ROI images are shown in panels~\ref{fig:supergranule_proc}.d's. To check for the detection accuracy visually, we have over-plotted the detected supergranular boundaries on top of the histogram equalized images as shown in panels~\ref{fig:supergranule_proc}.e's and find a very good match between the two. We also provided magnified 400"$\times$400" views of panels~\ref{fig:supergranule_proc}.e's in panels~\ref{fig:supergranule_proc}.f's to depict the overlap of network with watershed boundaries (shown in dark blue double lines).
After detecting the supergranules from every image using the above described method, we now define some of the parameters associated with it in the next section. 

   \subsection{Scale, Cricularity and Fractal Dimension of Supergranules}\label{sec:fractal}

               All the detected supergranules, from every single image, have been isolated using the region labelling method \citep{sonka2014image}. To calculate the `supergranulation scale' (characteristics scale of these structures), we equate each supergranule area to the area of a circle and the radius of this circle is defined as the `supergranule scale'. We take average of these radii for each image to find a number named as `Average Supergranule Scale'.

\begin{figure}[!htb]
\centering
  \includegraphics[width=0.8\linewidth]{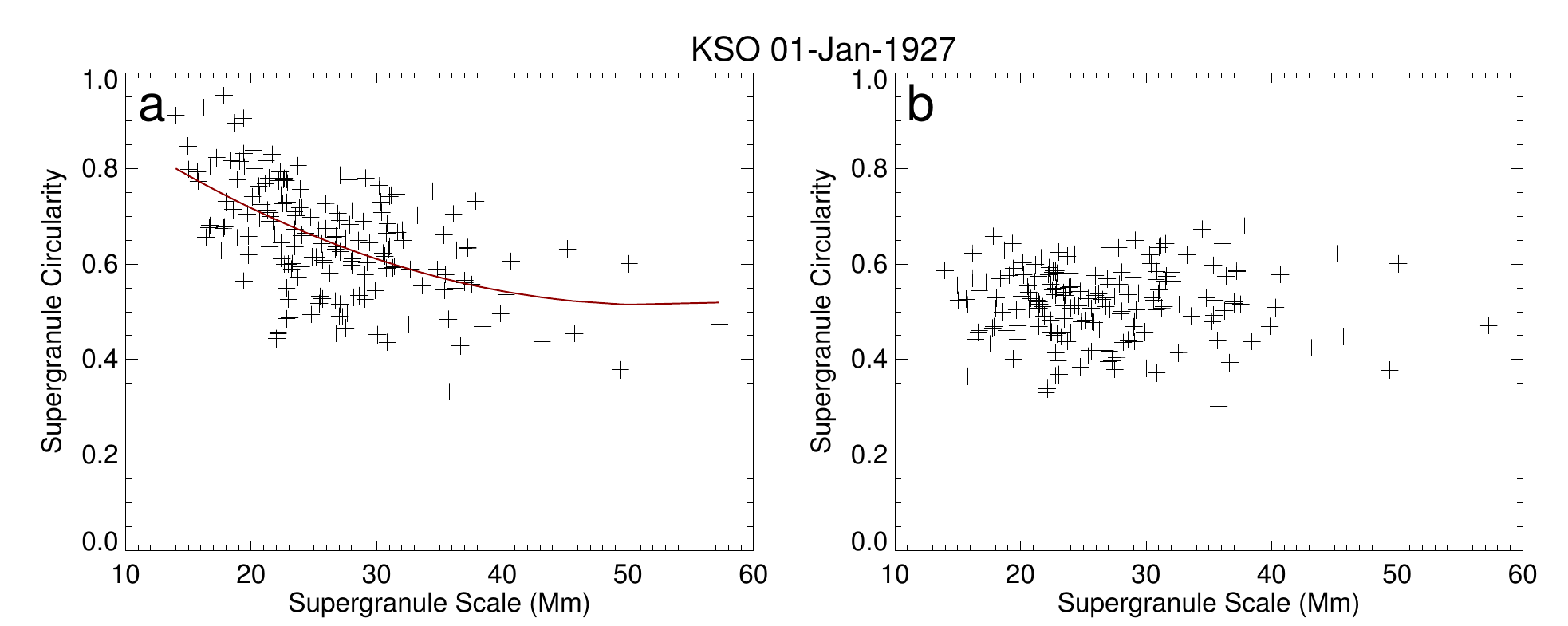}
  \caption{ a) Dependence of supergranule circularity with the scale. The solid red line is the fitted 2$^{nd}$ degree polynomial on the observed trend ; b) Same as previous only after removing the trend. }
 \label{fig:circ}
\end{figure}

Next, we define the circularity of each supergranule by the expression ${4\pi A}/{P^{2}}$ where A and P denotes area and perimeter of each supergranule \citep{0004-637X-534-2-1008}. In the digital domain, circularity shows some dependence on the size of the structures. This arises because the feature boundaries get exaggerated when size decreases (due to the fixed pixel resolution). To correct for this dependency, we first calculate a trend of circularity versus scale, for each image, by fitting a 2$^{nd}$ degree polynomial (Figure~\ref{fig:circ}.a) and then the data points are divided by the fitted curve to correct for the trend (Figure~\ref{fig:circ}.b). As larger areas are assumed to have correct (scale independent) circularity, we have multiplied the normalized circularity values by minimum of the polynomial curve. 

One of the other important parameter associated with a supergranule, is the fractal dimension which is a measure of complexity and self-similarity of a structure \citep{mandelbrot1982fractal}. It is also termed as fractional dimension and it captures the dependance of structure details on scale. Now, the fractal dimension (D) is estimated from the area (A) and perimeter (P) of a given structure, via the relation $P \propto A^{\frac{D}{2}}$ \citep{Muller1994}. Thus, twice the slope of the area vs perimeter plot, in a log-log scale, will be equal to the fractal dimension \citep{Paniveni11022010}.

   \section{Results From KSO} \label{sec:kresult}
   \subsection{Full ROI; The Aggregate}

  \begin{figure}[!htb]
  \centering
  \includegraphics[width=1\linewidth]{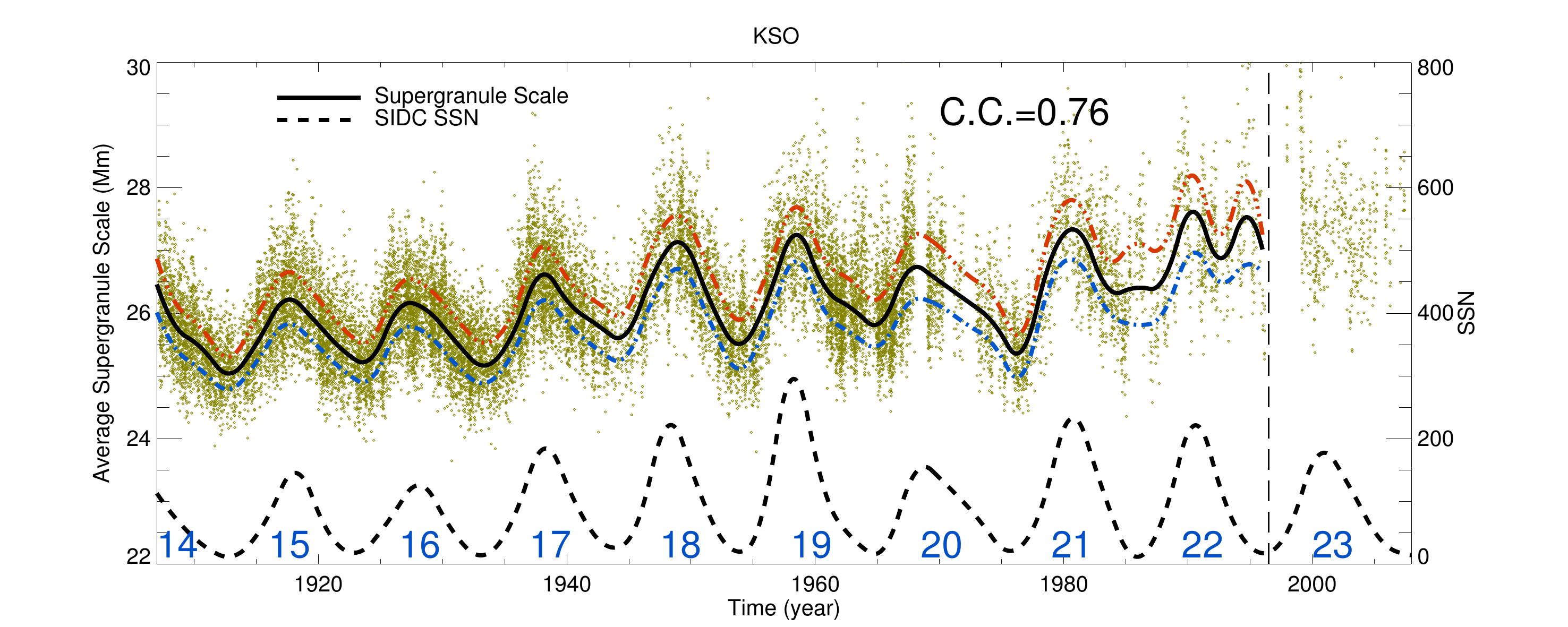}
  \caption{Cyclic variation of mean supergranule scale for 9 cycles. Green points correspond to data from KSO.  The black solid curve is smoothed average supergranule scale. Blue and red dashed curves depict spline smoothing for first and third quartiles respectively. The black dashed curve depicts temporal variation of smoothed sunspot number (SSN). Cycle numbers are marked in blue.}
 \label{fig:supergranule_scale}
\end{figure} 
As mentioned in the previous section, we have calculated different supergranule parameters like scale, circularity and the fractal dimension within a selected ROI (we call it `aggregate' hereafter) from each of the KSO C\MakeLowercase{a}  $\scriptsize{{\textrm{II}}}$ K images. Figure~\ref{fig:supergranule_scale} shows the variation of average supergranule scale over 9 cycles (cycle 14 to cycle 22) studied in this paper. The green dots correspond to average scales determined from individual images and the solid black curve represents smoothed version of the same. We have additionally plotted spline smoothed first and third quartile curves with blue and red dashes respectively in Figure~\ref{fig:supergranule_scale} and it shows that scatter in the data is less than the temporal variation of the parameter.

All the smoothed curves presented in this paper are generated using the CRAN package named `cobs' of the statistical analysis software R (details about this can be found in \citet{feigelson2012modern}). This smoothing technique is based on basis splines \citep{Reinsch:1967:SSF:2715402.2715725} and allows manual input features such as constraints and knot points. For this study we have used a quadratic spline with the penalty parameter $\lambda$ \citep{hastie1990generalized} set to 1. This method takes care of the temporal variation of the data spread (or sudden discontinuities) and is effective in avoiding the `artificial' jitters as opposed to the conventional running average technique. Moreover, to compare the two methods (spline smoothing and the running averaging) we have repeated all of the presented analysis with a running averaged data and the results are presented in Appendix.

The vertical dashed line in Figure~\ref{fig:supergranule_scale} indicates that the KSO image quality degraded substantially after this period (1997 onwards/cycle 23 onwards). Though we have detected and calculated all the supergranule parameters using the images from this time period also (1997 onwards, as shown in the plots), but all the correlation values (with the SSN) have been calculated for a period 1907-1996. To extend our analysis beyond 1996, we used the PSPT-Italy (1996-2016), PSPT-USA (2005-2015) data to cover the rest period and the results from them are discussed in subsequent sections.

Now from the curve we immediately notice that it has sunspot cycle like periodicity of $\sim$11 years. To understand its connection with the solar cycle more clearly, we have over-plotted the smoothed sunspot number (SSN) data in the same panel (dashed curve). A positive correlation value of 0.76 confirm the in-phase variation of the average supergranule scale with the sunspot number. The calculated scale values (Figure~\ref{fig:supergranule_scale}) vary from 24 $\mathrm{Mm}$ (during the cycle minima) to 28-30 $\mathrm{Mm}$ (during the cycle maxima) with an average around 26 $\mathrm{Mm}$. These estimated scale values from the KSO data match closely with the same presented in \citet{2041-8205-730-1-L3} where the authors have used Mount Wilson Solar Observatory (MWO) historical data for three cycles (1944-1976). It must be emphasized here that \citet{2041-8205-730-1-L3} could not get a clear trend of the in-phase variation of supergranule radius with the SSN in all the three cycles they analyzed. In our analysis, if we look for the same period as presented in \citet{2041-8205-730-1-L3} i.e from 1944 to 1976, we notice that the in-phase variation signature is prominently visible for all of the cycles. In fact, the one to one correlation with the SSN is clearly demonstrated for all the cycles (cycle 14-22) investigated in this study. Thus, we conclude that the long-term data availability at KSO has enabled us to establish the in-phase variation of the radius parameter with SSN over much greater span of time than any other previous studies.

\subsection{Active And Quiet Regions}

Since the major magnetic activities are concentrated on the active regions (ARs), it would be interesting to investigate the effect of the same on the different properties of supergranules in ARs and on the rest of the Sun, the quite regions (QRs).

\begin{figure}[!htbp]
  \centering
  \includegraphics[width=0.8\linewidth]{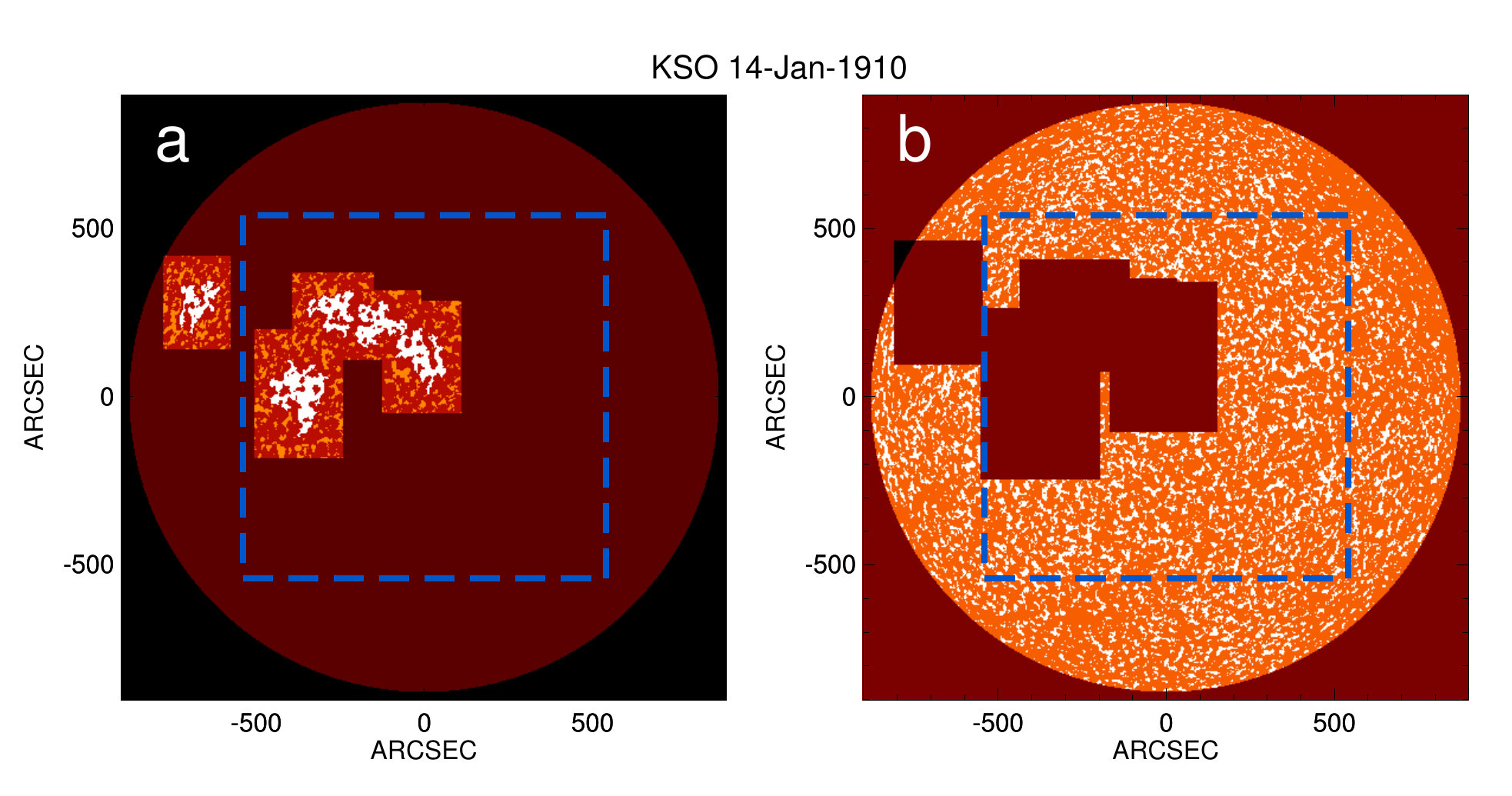}
  \caption{Separation of active and quiet region supergranules. a) Rectangular regions with white plage structures are defined as Active Regions (ARs); b) Regions away from plage structures (outside the rectangular patches and inside the disc) are considered as Quiet Regions (QRs). Dashed rectangles, in both the panels, show the regions within which supergranule detection is performed (this is same as the ROI). }
 \label{fig:supergranule_reg}
\end{figure}

 Though there have been attempts to divide the solar images into ARs and QRs and study the changes in supergranule scale, all of them were for a very short span of time, at most for one solar cycle \citep{{1989A&A...213..431M},{1999A&A...344..965B}, {2008A&A...488.1109M}}. In this study we have recorded the supergranule parameters, separately for AR and QR, for more than 9 cycles with a fully-automated method.
To identify the locations of the ARs from the C\MakeLowercase{a}  $\scriptsize{{\textrm{II}}}$ K images, we have used the plage locations as proxies for the magnetic field \citep{2011ApJ...730...51S}. All the full disc limb darkening corrected KSO C\MakeLowercase{a}  $\scriptsize{{\textrm{II}}}$ K images were used to detect plages with a fully-automated method as described in \citet{0004-637X-827-1-87}. Next we used a rectangular mask, around each of the detected plage, with sides 3 times the maximum distances of plage structure coordinates from centroid along $\mathrm{X}$ and $\mathrm{Y}$. We define such rectangular regions as ARs. This procedure is shown, for a representative KSO image, in Figure~\ref{fig:supergranule_reg}.a. We keep a margin of 0.5 times of those $\mathrm{X}$ and $\mathrm{Y}$ distances and region beyond that margin is considered as QR (Figure~\ref{fig:supergranule_reg}.b). Supergranule detection was performed within a rectangular region about the disc center as shown in Figures{~\ref{fig:supergranule_proc},~\ref{fig:supergranule_reg}}.

  \begin{figure}[!htbp]
  \centering
  \includegraphics[width=0.85\linewidth]{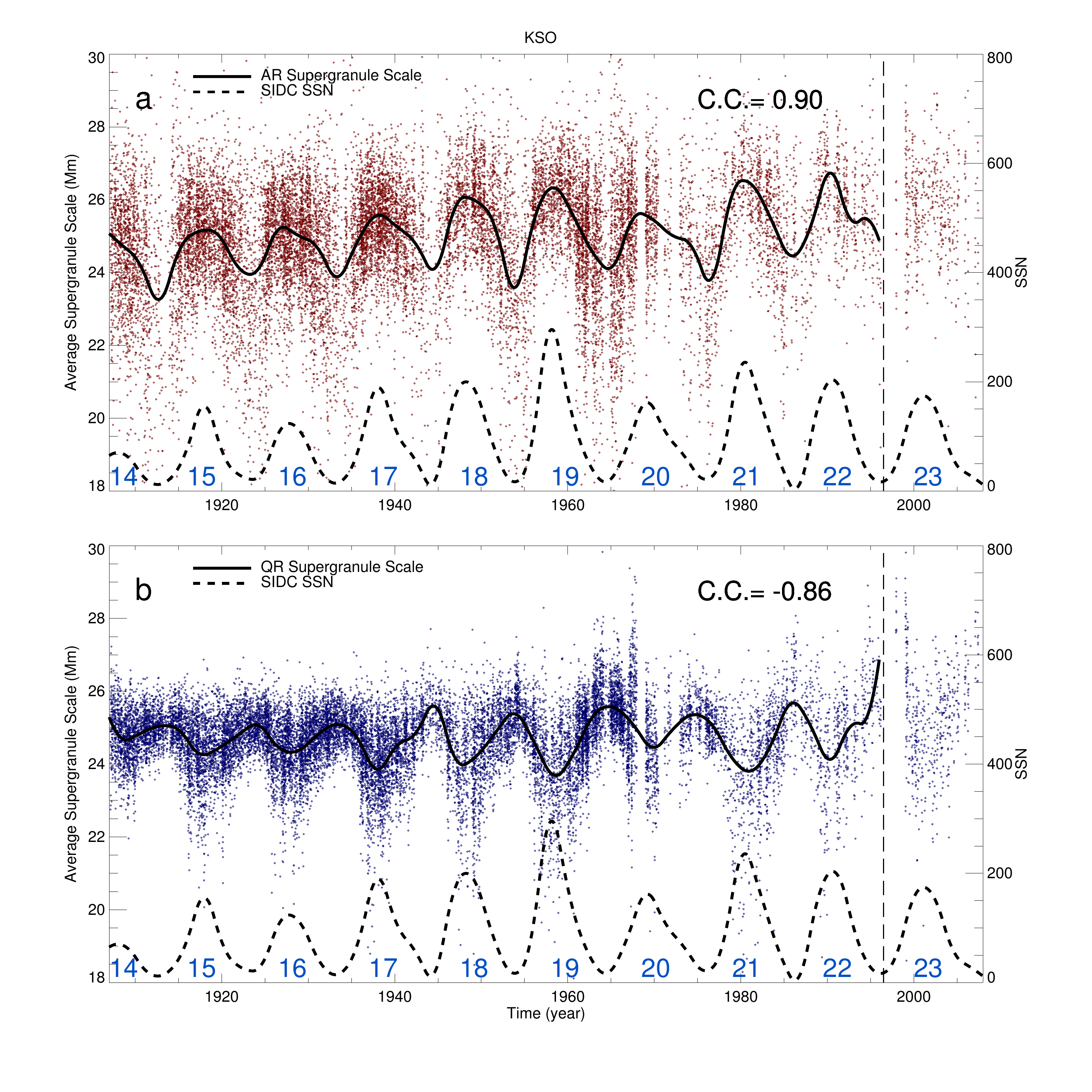}
  \vspace{-.05\textwidth}
  \caption{Cyclic variation of mean supergranule scale for the AR (panel a) and for the QR (panel b). Dots correspond to the measured mean scale values from individual images whereas the solid curves represent the smoothed versions of the same. The SSN cycle is shown with a dashed curve. Individual Cycle numbers are marked in blue. The color theme i.e. results from ARs in red and results from QRs in blue, is followed throughout this paper.}
  \label{fig:supergranule_scale_act_qt}
\end{figure}

 Figure~\ref{fig:supergranule_scale_act_qt} shows the temporal variation of AR and QR supergranule scales with solar cycles. The AR supergranular scale varies coherently with the SSN (plotted as black dashed curve). The correlation coefficient between the two equals to 0.90. Apart from this in-phase variation with the sunspot cycle, we notice that the average AR scale, in this case, is around 25 $\mathrm{Mm}$ (we obtained similar number from the aggregate case as shown in Figure~\ref{fig:supergranule_scale}). The temporal variation of QR supergranule average scale is illustrated in Figure~\ref{fig:supergranule_scale_act_qt}.b. Interestingly, for the QR case we find a strong anticorrelation between the mean scale and the SSN cycle. The correlation coefficient is -0.86. For both the AR and the QR, we find substantial cases when the scale values have comparatively low numbers ($\approx$22 $\mathrm{Mm}$).

  \begin{figure}[!htbp]
  \centering
  \includegraphics[width=1\linewidth]{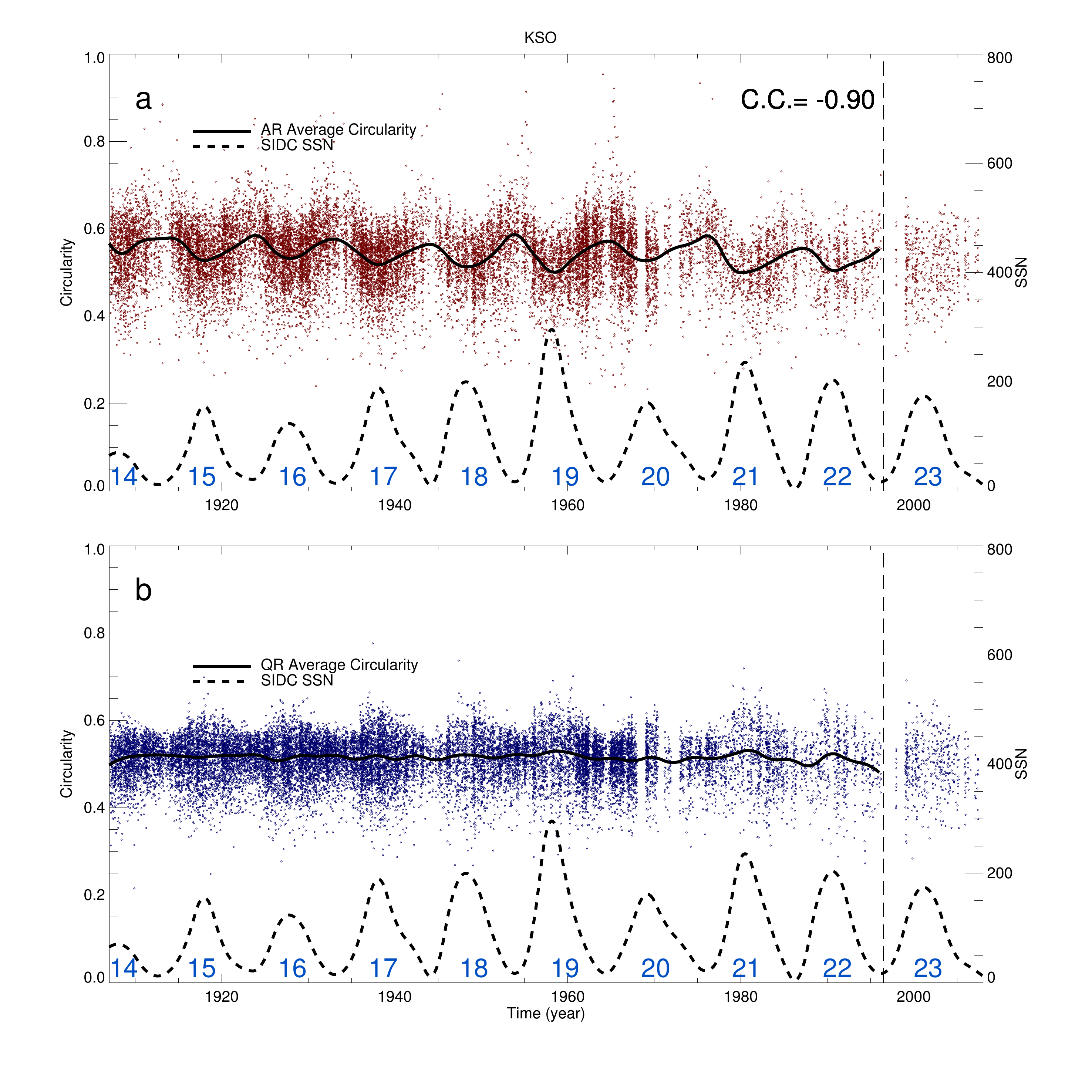}
  \vspace{-.1\textwidth}
  \caption{Variation of supergranule circularity for over a century for ARs (panel a) and for QRs (panel b). Dots correspond to circularity values obtained from individual images whereas the solid curves represent their smoothed version. The SSN cycle is also plotted at the bottom of each panel.}
 \label{fig:supergranule_circ_qt_act}
\end{figure}

We have also calculated the scale normalised average circularity, separately for AR and QR, and the results are shown in panels a and b of Figure~\ref{fig:supergranule_circ_qt_act}, respectively. From the evolution of the AR circularity, we observe that the supergranules are more circular during the solar minima as opposed to the solar maxima. A correlation coefficient of -0.90 confirms the same. For the QR (panel~\ref{fig:supergranule_circ_qt_act}.b), it becomes interesting as the circularity parameter shows no apparent correlation with the sunspot cycle.

\begin{figure*}[!htbp]
  \centering
  \includegraphics[width=0.9\linewidth]{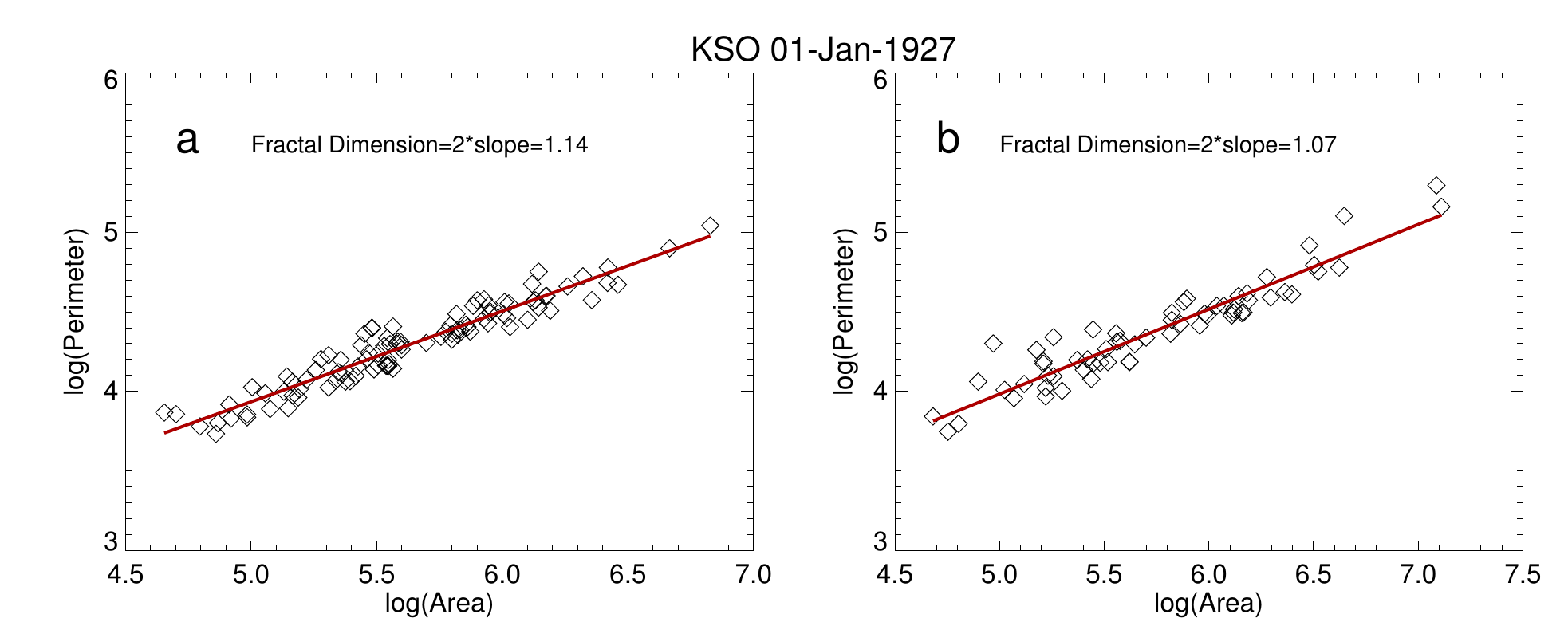}
  \caption{Calculation of fractal dimension of supergranules. a) Fractal dimension of supergranules in QRs; b) Fractal dimension of supergranules in ARs.}
 \label{fig:fract}
\end{figure*}

Next we calculate the fractal dimension for the ARs and the QRs. As defined in Section~\ref{sec:fractal}, the fractal dimension is equal to the twice of the slope of log-log area vs perimeter plot. Different panels of Figure~\ref{fig:fract} show the calculation of the fractal dimension for the AR and the QR from a single KSO image. Previously, \citet{Paniveni11022010} have quantified the AR and QR fractal dimension of supergranules (identified manually) from KSO Ca~{\sc ii}~K filtergrams  for  period of 1.5 years between 2001 and 2002. In this study we have recorded the same for a much longer period and also using a fully automated method.

\begin{figure}[!htbp]
 \centering
  \includegraphics[width=0.95\linewidth]{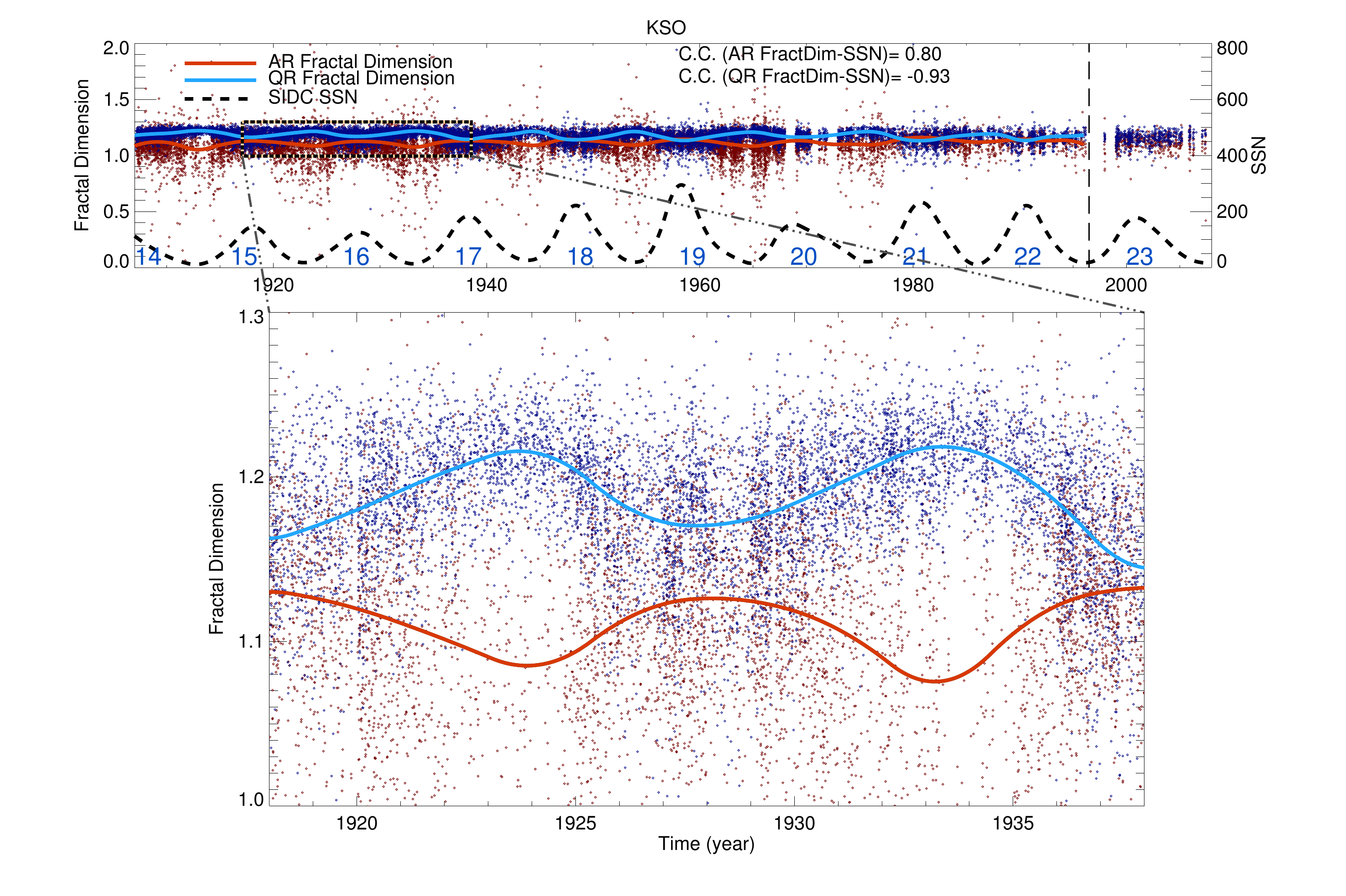}
  \caption{Variation of supergranule fractal dimensions corresponding to individual images in ARs (red dots) and QRs (blue dots). Smoothed curves for the two are shown in solid red and pale-blue curves respectively. Bottom panel presents the magnified view of the region enclosed by dotted rectangle in the top panel.}
 \label{fig:supergranule_fract_qt_act}
\end{figure}

 Figure~\ref{fig:supergranule_fract_qt_act} shows the temporal variation of the fractal dimension for the two regions as obtained from the KSO data. For the AR fractal dimension we observe a good correlation (with correlation coefficient 0.80) with the solar cycle. For the QR, it is exactly opposite i.e the QR fractal dimension has a strong anti-correlation (with correlation coefficient -0.93) with the solar cycle. Additionally we notice that fractal dimension for active regions is lower than the same for quiet regions on an average as the smoothed pale-blue curve goes rarely below smoothed red curve (in accordance with \citet{Paniveni11022010}).

\section{Results From PSPT} \label{sec:presult}

As mentioned in the previous section, the KSO data quality degraded after 1996 and thus the calculated supergranule parameters have more scatter, less data points and large discontinuities.

 \begin{figure*}[!htbp]
  \centering
  \includegraphics[width=.75\linewidth]{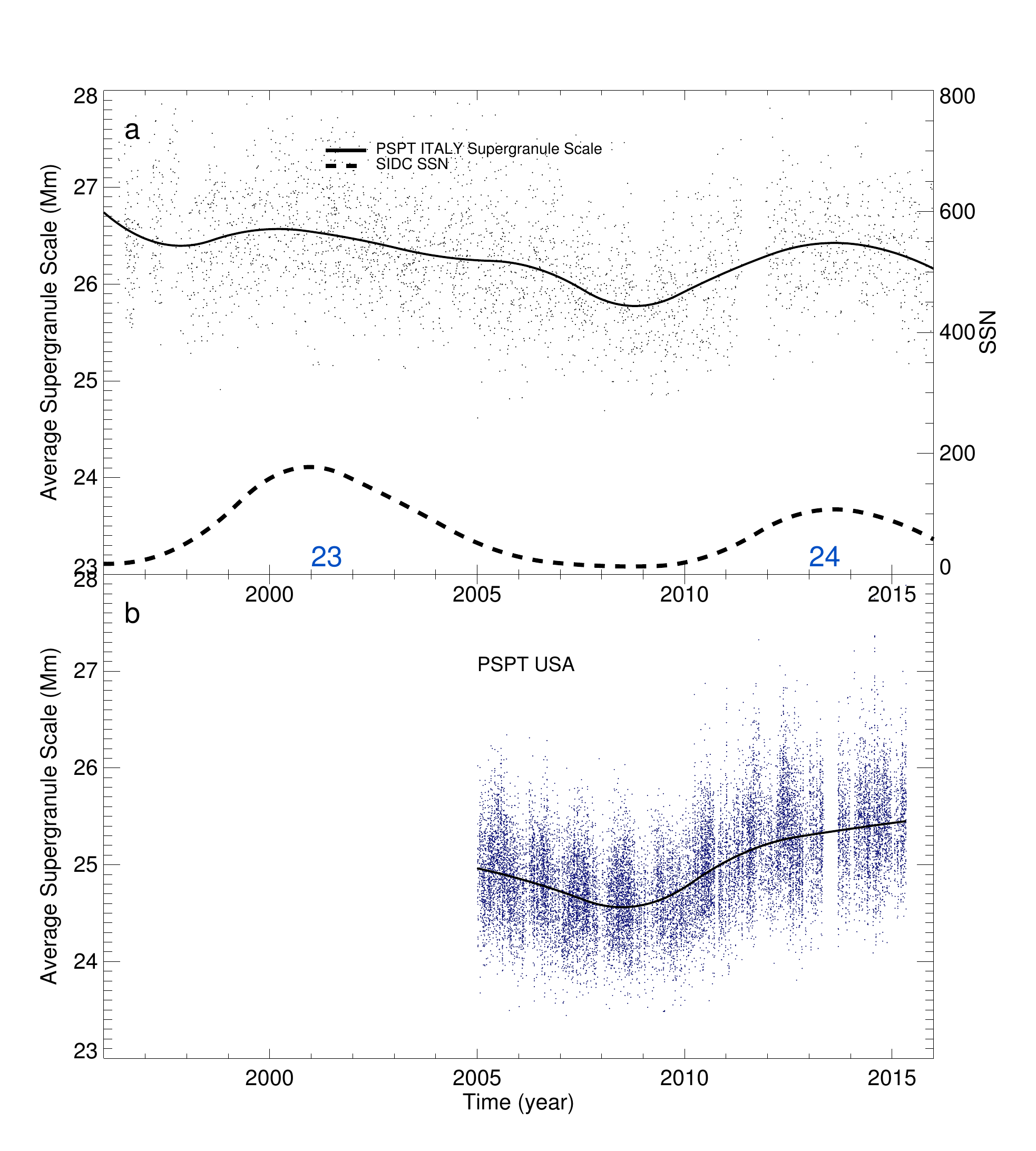}
  \vspace{-.04\textwidth} 
  \caption{Cyclic variation of mean supergranule scale for 1996-2016 from various observatories.  }
 \label{fig:supergranule_vari_obs}
\end{figure*} 

 We, have thus used the same technique (of supergranule detection) on the PSPT-Italy and PSPT-USA images. Results from these data sets are shown in different panels of Figure~\ref{fig:supergranule_vari_obs}. From the plot we note that the average supergranule scales from two observations match well with each other. In fact, they are also close to the value obtained from the KSO data (Figure~\ref{fig:supergranule_scale}). Upon careful observation of the plot we find some differences though. The plot reveals that the there is a constant shift in the measured supergranule scale values ($\approx$1 $\mathrm{Mm}$) between the PSPT-Italy and PSPT-USA data. The reason behind this may be the fact the PSPT Italy data are available in `JPEG' format which introduces some compression in the original image whereas PSPT-USA data are available in standard astronomical FITS format without any compression.

In panel~\ref{fig:supergranule_vari_obs}.a we notice that the change in the mean scale value , from PSPT-Italy, does not show a clear in-phase variation with the cycle 23 (1996-2008). In the next cycle, cycle 24 (2008 onwards), we revive the solar cycle like variation. For PSPT-USA (panels ~\ref{fig:supergranule_vari_obs}.b), we do find an in-phase variation of the same with solar cycle. Thus we conclude that the results from the PSPT-Italy is not due to an detection artefact but rather related to the image quality (or continuity) of the telescope.

 \section{Summary and Conclusion} \label{sec:summary}

 In this study, we have used. for the first time, the calibrated C\MakeLowercase{a}  $\scriptsize{{\textrm{II}}}$ K images recorded from KSO to identify different supergranule parameters such as mean radius, circularity, fractal dimension for a century (1907-2007). This has been the longest time series till date for the supergranule geometrical parameters. The main findings are listed below:

$\bullet$ We have implemented a fully automated algorithm to detect the supergranules from the intensity images. Using this automated segmentation method  we find the mean supergranule scale to vary between a range of 22 Mm to 28 Mm which is similar to the one presented in \citet{2041-8205-730-1-L3}.

$\bullet$ To isolate the effect of strong, large-scale magnetic field on the derived supergranule parameters, we segregate the ARs and QRs from every intensity image using the plages as proxies for the magnetic locations. Analysis shows that the AR supergranule mean scale varies in phase with the solar cycle whereas for the QR supergranule mean scale it is the opposite. AR supergranule scale fluctuation about mean is more than that of QR values. So, though QR scale is anticorrelated with SSN, AR scale dominates to make aggregate scale in phase with SSN. We conjecture that, AR scale fluctuation is influenced by the spatial extent of magnetic field. It other words, bigger active regions during solar maxima might be causing the AR supergranules to become bigger. Network magnetic elements have shrinking effect on supergranules as hinted by \citet{2008A&A...488.1109M}. During minima those reduce and might cause larger QR supergranules . 

$\bullet$ The circularity parameter is found to behave differently for the two regions (ARs and QRs). AR circularity shows a strong anti-correlation with the sunspot cycle whereas the QR circularity shows no dependence. It may be that the `random walk' associated with network magnetic element causes the active region supergranules  to distort and become less circular during cycle maxima.

$\bullet$ Fractal dimension, the measure of the boundary irregularity, also has different evolution for the two regions. In this case, the AR fractal dimension has a positive correlation with the sunspot cycle whereas the QR has a negative correlation.

$\bullet$ We also used our detection technique on different data sets from different observatories. The similar parameter values clearly depicts the robustness of the technique used in this paper.

In conclusion, we have used a unique data set to study the variation of the supergranular parameters with the solar cycle. The variation of  supergranule parameters also have an effect on the `total solar irradiance' \citep{2041-8205-730-1-L3}. In our future work we would like to explore further on this topic using data from different observatories. Now, active regions are the locations of strong large-scale magnetic fields (mostly the sunspot fields) which are believed to be generated by the global solar dynamo \citep{lrsp-2010-3} whereas the small scale quiet Sun magnetic field is believed to be governed by a local dynamo \citep{2012A&A...547A..93S}. The different nature of the correlations for AR and QR supergranules thus, reflect this inherent difference in the nature of the magnetic fields.  It is not clear how magnetic field is influencing scale variability but our results of segregating the AR and QR do provide new constraints that we hope future magneto convection models will be able to explain.

  \section{acknowledgements}
  We would like to thank the Kodaikanal facility of Indian Institute of Astrophysics, Bangalore, India for providing the data. These data are now available for public use at  \url{https://kso.iiap.res.in/data}.  We also thank the SCIENCE \& ENGINEERING RESEARCH BOARD (SERB) for the project grant (EMR/2014/000626).  
 
  \section{Appendix} \label{sec:append}
  
\begin{figure*}[!htbp]
\centering
 \captionsetup[subfigure]{labelformat=empty}
  {\includegraphics[trim = 0mm 0mm 0mm 0mm, clip,width=\textwidth]{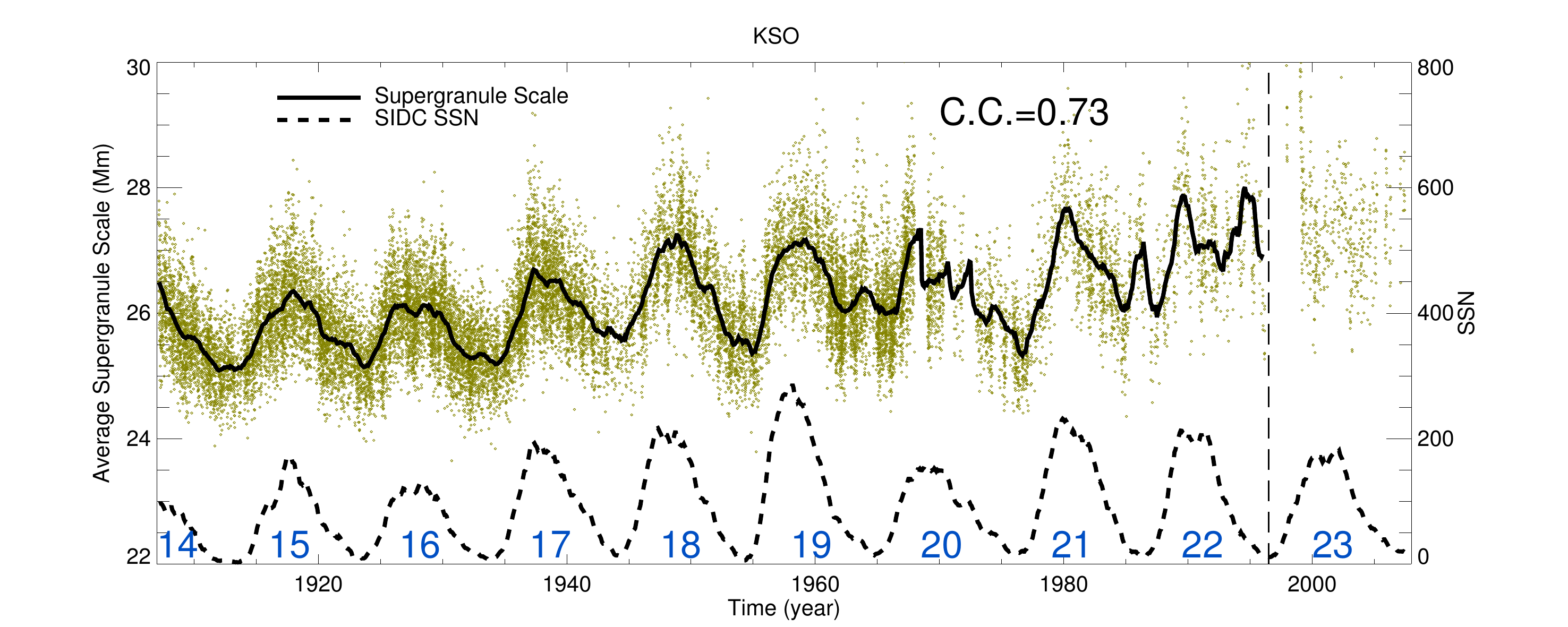}}
  {\includegraphics[trim = 0mm 0mm 0mm 0mm, clip,width=\textwidth]{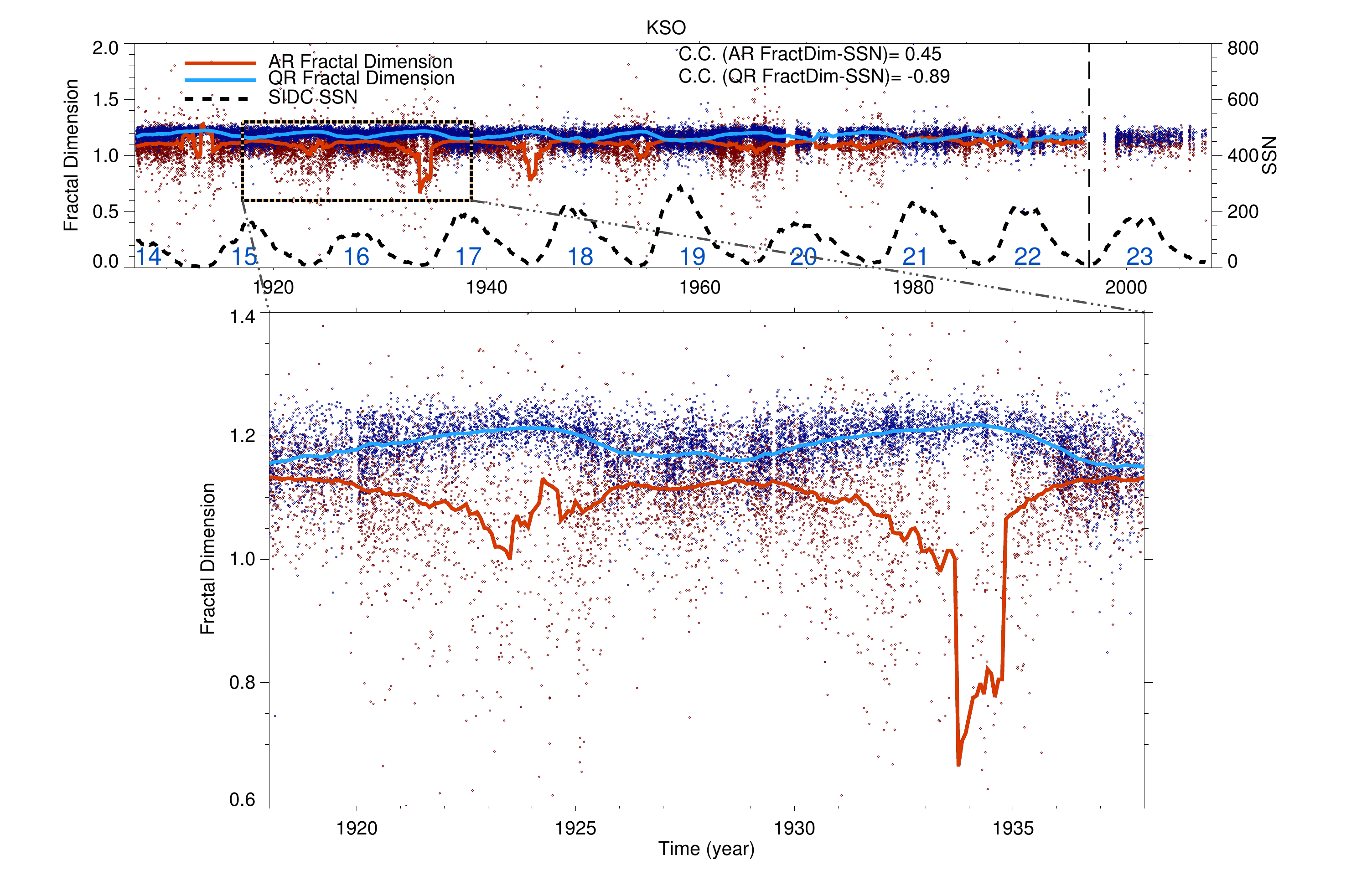}}
  \caption{Cyclic variation of supergranule parameters for 9 cycles. The upper panel shows aggregate scale variation and the bottom panel shows cyclic variation of fractal dimension for AR and QR. These figures are same as Figure~\ref{fig:supergranule_scale} and Figure~\ref{fig:supergranule_fract_qt_act} except the smoothing approach being running averaging.}
 \label{fig:fig_append}
\end{figure*}

  As mentioned earlier, we have re-computed all the correlation coefficients with running averaged curves in order to check for the robustness of the obtained results. Two of such plots have been shown in Figure~\ref{fig:fig_append}. The top and the bottom panels in this plot are similar to the 
Figures{~\ref{fig:supergranule_scale} and~\ref{fig:supergranule_fract_qt_act}} with the smoothing technique being running averaging. Comparing the respective plots, we observe that the running averaging generates are much jittery curves compared to the spline smoothed ones. Such jitters actually results in a slightly lower correlation coefficient (C.C.) values. However, the improvement in correlations are mostly marginal except for one case (SSN vs AR fractal dimension). The C.C.s obtained from these two methods are listed in Table~\ref{table:correl}. Thus, the closeness of the correlation values confirm the fact that the physical interpretations are not affected by the smoothing methods.
  
 \begin{table}[!htbp]
\begin{center}
\centering
\caption{Comparison of correlation coefficients (C.C.)  for two different smoothing techniques}  

\label{table:correl}
\begin{tabular}{lcccc r@{   }l c} 

  \hline
     & \multicolumn{2}{r}{C.C.}\\
     & \multicolumn{1}{c}{Correlated Data pair}& \multicolumn{1}{c}{Running average} & \multicolumn{1}{c}{Spline smoothing} \\
      \hline
     &  SSN - Aggregate scale & 0.73 & 0.76 \\
     &  SSN - AR scale & 0.86 & 0.90 \\
     &  SSN - QR scale & -0.81 & -0.86 \\
     &  SSN - AR circularity & -0.82 & -0.90 \\
     &  SSN - AR fractal dimension & 0.45 & 0.80 \\
     &  SSN - QR fractal dimension & -0.89 & -0.93 \\
      \hline

\end{tabular}
\end{center}
\end{table}

 \bibliographystyle{apj}

 \end{document}